\documentclass[twocolumn,showpacs,preprintnumbers,amsmath,amssymb,prb,
superscriptaddress]{revtex4}

\usepackage{graphicx}
\usepackage{dcolumn}
\usepackage{bm}

\begin{document}

\preprint{}

\title{Comment on ``Wave-scattering formalism for thermal conductance in thin wires with surface disorder''}

\newcommand{\dilog}{\mathrm{dilog}}

\author{Marcos G. Menezes}
 \email{marcosgm@if.ufrj.br}
\affiliation{Instituto de F\'{i}sica, Universidade Federal do Rio de Janeiro, Rio de Janeiro, RJ, Caixa Postal 68528 21941-972, Brazil}

\author{Jordan Del Nero}
\affiliation{Instituto de F\'{i}sica, Universidade Federal do Rio de Janeiro, Rio de Janeiro, RJ, Caixa Postal 68528 21941-972, Brazil}
\affiliation{Departamento de F\'{i}sica, Universidade Federal do Par\'{a}, Bel\'{e}m, PA, 66075-110, Brazil}

\author{Rodrigo B. Capaz}
\affiliation{Instituto de F\'{i}sica, Universidade Federal do Rio de Janeiro, Rio de Janeiro, RJ, Caixa Postal 68528 21941-972, Brazil}

\author{Luis G. C. Rego}
\affiliation{Departamento de F\'{i}sica, Universidade Federal de Santa Catarina, Florian\'{o}polis, SC, 88040-900, Brazil}

\date{\today}

\begin{abstract}
In their calculations based on the Landauer transport equation, Akguc and Gong \lbrack Phys. Rev. B 80, 195408 (2009)\rbrack \space obtained an expression for the heat conductance of a quantum wire valid in the ballistic regime and in the limit of vanishing temperature difference between reservoirs. Their result appears to be different from the one reported in the previous paper of Rego and Kirczenow \lbrack Phys. Rev. Lett. 81, 232 (1998)\rbrack , which led them to argue that their new result was the correct one. We show here that, in fact, both results are correct, since different definitions for the dilogarithm function were used in those papers. Hence, comparisons between these two results should be done with care.
\end{abstract}

\pacs{73.23.Ad, 65.90.+i}
\maketitle

In order to obtain an expression for the heat conductance of a quantum wire between reservoirs with different temperatures in the ballistic regime, both papers of Akguc and Gong (AG)\cite{gong} and Rego and Kirczenow (RK)\cite{rego} have a common starting point: The Landauer transport equation for the heat flux
\begin{equation}
\dot{Q} = \sum_\alpha \int_{0}^{\infty} \frac{dk}{2\pi} \hbar\omega_\alpha(k)v_\alpha(k)(\eta_R - \eta_L)\zeta_\alpha(k),
\end{equation}
where $\hbar\omega_\alpha(k)$ is the phonon energy and $v_\alpha(k)$ its velocity, $\zeta_\alpha(k)$ is a transmission coefficient and $\eta_i(\omega) = \lbrack e^{\hbar\omega \slash k_B T_i} - 1 \rbrack^{-1}$, $i = L, R$ is the phonon occupation number in each reservoir. Setting $\zeta_\alpha(k) = 1$ in the ballistic regime and using $\kappa = \dot{Q} \slash \Delta T$ for the conductance, with $\Delta T = T_R - T_L$, RK obtained in the limit $\Delta T \to 0$ the expression
\begin{eqnarray}
\label{krego}
\frac{\kappa}{\frac{k_B^2}{h} T} &=& \frac{\pi^2}{3}N_\alpha + \nonumber \\ &+& \sum_{\alpha '}^{N_{\alpha '}} \left\{ \frac{\pi^2}{3} + 2 \dilog(e^{x_0}) + \frac{x_0^2 e^{x_0}}{e^{x_0} - 1} \right\},
\end{eqnarray}
where $N_\alpha$ is the number of modes with zero cutoff frequency (massless modes), $N_{\alpha '}$ is the number of modes with non-zero cutoff frequency $\omega_{\alpha '}(0)$ and $x_0 = \hbar\omega_{\alpha '}(0) \slash k_B T$ with $T = (T_R+T_L)/2$ the average temperature of the reservoirs. For a single-mode with non-zero cutoff frequency, this expression becomes
\begin{eqnarray}
\label{krego2}
\frac{\kappa}{\frac{k_B^2}{h} T} =  \frac{\pi^2}{3} + 2 \dilog(e^{x_0}) + \frac{x_0^2 e^{x_0}}{e^{x_0} - 1}.
\end{eqnarray}
Here and in Eq.(\ref{krego}), the function $\dilog(z)$, known as the dilogarithm function is defined as the integral\cite{handbook}
\begin{eqnarray}
\label{dilog}
\dilog(z) = \int_1^z \frac{\ln t}{1-t} dt.
\end{eqnarray}
This definition associated with the ``$\dilog$'' notation is the one adopted for example in the software Maple\cite{maple}. Note that $\dilog(e^{x_0})$ is a real function for all values of $x_0$ and since $\dilog(1) = 0$ for $x_0 = 0$, the correct limit of high temperatures or zero cutoff frequency is obtained.

On the other hand, by first taking the limit $\Delta T \to 0$ on the expression for $\kappa$ and then evaluating the resulting integral, AG obtained the result for a single mode with non-zero cutoff frequency as
\begin{eqnarray}
\label{kgong}
\frac{\kappa}{\frac{k_B^2}{h} T} &=& \frac{x_0^2 e^{x_0}}{e^{x_0} - 1} - \nonumber \\ &-& 2 x_0 \ln (1 - e^{x_0}) + \frac{2\pi^2}{3} - 2 Li_2(e^{x_0}) .
\end{eqnarray}
Here, the function $Li_2(z)$ is also known as the dilogarithm function, but when used with this notation, it is defined in a slightly different form as\cite{lewin}
\begin{eqnarray}
\label{polylog}
Li_2(z) = \int_z^0 \frac{\ln (1-t)}{t} dt.
\end{eqnarray}
This definition and notation for the dilogarithm function is the one adopted for example in the Mathematica\cite{mathematica} software. It is associated with the more general polylogarithm function $Li_n(z)$, defined as a power series\cite{lewin} that has the integral representation of Eq.(\ref{polylog}) for $n=2$. This function is different from the $\dilog(z)$ function defined earlier. In fact, from Eq.(\ref{dilog}) we obtain an important relation between the definitions
\begin{eqnarray}
\label{relacao}
\dilog(z) = Li_2(1 - z).
\end{eqnarray}
Note that $Li_2(e^{x_0})$ is a complex function, but its imaginary part is cancelled by the one of $\ln (1 - e^{x_0})$ in Eq.(\ref{kgong}), so we can drop the real part on $Li_2$ of Eq.(16) of AG's paper by flipping their log argument, as we did above. Note also that $Li_2(1) = \pi^2 / 6$ for $x_0 = 0$, so AG's expression Eq.(\ref{kgong}) also satisfies the correct limit of high temperatures or zero cutoff frequency. However, AG noted the ``clear'' difference between Equations \ref{krego2} and \ref{kgong} and stated that RK's expression, Eq.(\ref{krego2}), was wrong. They pointed out some differences between the expressions, but apparently they weren't aware that the $\dilog(z)$ function in RK's paper isn't the same function as the $Li_2(z)$ they used. This fact led them to incorrect conclusions, such as the one of Fig. 4b in their article, where RK's expression is plotted after the wrong substitution $\dilog(z) = Li_2(z)$ and therefore a physically incorrect result is obtained, where the conductance doesn't even go to zero as $T$ goes to zero.

In fact, there is a ``reflection'' property of the $Li_2(z)$ function\cite{lewin}
\begin{eqnarray}
\label{reflexao}
Li_2(z) + Li_2(1 - z) = \frac{\pi^2}{6} - \ln(z)\ln(1 - z)
\end{eqnarray}
that allows us to estabilish a relation between the expressions of RK and AG. By setting $z = e^{x_0}$ and using Eq.(\ref{relacao}), we can see that RK's expression leads to Akguc's expression and vice-versa, so both results are correct when compatible definitions are considered, as should be expected from the high temperature or zero cutoff limits we discussed above.

Finally, we remark that extra care must be taken when referring to ``the'' dilogarithm function. In fact, due to historical reasons, there are several definitions for this function in the literature\cite{lewin} other than the two we discussed here. Therefore, authors should be aware of that when comparing their results with others in the literature. We emphasize, however, that the comparison between RK and AG represents a small portion of AG's article and most of the results of the latter paper are sound and important.

\begin{acknowledgments}
We acknowledge the financial support from the brazilian funding agencies: CAPES, CNPq, FAPERJ, FAPESPA and CNPq/INCT Nanomateriais de Carbono.
\end{acknowledgments}

\bibliography{commentgong}

\end{document}